# Tunneling magnetoresistance in trilayer structures composed of group-IV ferromagnetic semiconductor $Ge_{1-x}Fe_x$, MgO, and Fe

Yuki K. Wakabayashi,[1,a)] Kohei Okamoto,[1] Yoshisuke Ban,[1] Shoichi Sato,[1] Masaaki Tanaka,[1,2,b)] and Shinobu Ohya[1,2,c)]

[1]*Department of Electrical Engineering and Information Systems, The University of Tokyo, 7-3-1 Hongo, Bunkyo-ku, Tokyo 113-8656, Japan*

[2]*Center for Spintronics Research Network (CSRN), Graduate School of Engineering, The University of Tokyo, 7-3-1 Hongo, Bunkyo-ku, Tokyo 113-8656, Japan*

Abstract

Group-IV-based ferromagnetic semiconductor $Ge_{1-x}Fe_x$ (GeFe) is one of the most promising materials for efficient spin injectors and detectors for Si and Ge. Recent first principles calculations (Sakamoto *et al.*, Ref. 9) suggested that the Fermi level is located in two overlapping largely spin-polarized bands formed in the bandgap of GeFe; spin-down $d(e)$ band and spin-up $p$-$d(t_2)$ band. Thus, it is important to clarify how these bands contribute to spin injection and detection. In this study, we show the first successful observation of the tunneling magnetoresistance (TMR) in magnetic tunnel junctions (MTJs) containing a group-IV ferromagnetic semiconductor, that is, in MTJs composed of epitaxially grown $Fe/MgO/Ge_{0.935}Fe_{0.065}$. We find that the $p$-$d(t_2)$ band in GeFe is mainly responsible for the tunneling transport. Although the obtained TMR ratio is small (~0.3%), the TMR ratio is expected to be enhanced by suppressing leak current through amorphous-like crystal domains observed in MgO.

[a)]Electronic mail: wakabayashi@cryst.t.u-tokyo.ac.jp
[b)]Electronic mail: masaaki@ee.t.u-tokyo.ac.jp
[c)]Electronic mail: ohya@cryst.t.u-tokyo.ac.jp

Group-IV-based ferromagnetic semiconductor (FMS) $Ge_{1-x}Fe_x$ (GeFe) is one of the most promising materials for efficient spin injectors and detectors for Si and Ge without the conductivity mismatch problem,[1] because it can be epitaxially grown on Si and Ge substrates with flat and smooth interfaces and because its conductivity can be widely controlled by B doping.[2-5] The Curie temperature ($T_C$) of GeFe can be increased to 210 K,[6] which is higher than the highest $T_C$ value reported for (Ga,Mn)As (200 K).[7] Furthermore, X-ray magnetic circular dichroism measurements have revealed that nanoscale local ferromagnetic regions, which are formed in the local high-Fe-content regions, exist even at room temperature well above $T_C$.[8] Thus, GeFe has a potential for device applications operating at room temperature. Recently, first principles calculations suggested that the Fermi level ($E_F$) is located in two overlapping highly spin-polarized bands formed in the bandgap of GeFe; spin-down $d(e)$ band and spin-up $p$-$d(t_2)$ band.[9] Thus, it is important to clarify how these bands contribute to spin injection and detection. Thus far, there has been no report of successful detection of spin-dependent tunneling in magnetic tunnel junctions (MTJs) using group-IV FMSs. In this study, we have observed tunneling magnetoresistance (TMR)[10-13] in epitaxially grown MTJs composed of Fe/MgO/$Ge_{0.935}Fe_{0.065}$. This is the first observation of TMR in MTJs with a group-IV FMS. We found that spin-polarized carriers in the $p$-$d(t_2)$ band of GeFe are mainly responsible for the tunneling transport.

We fabricated MTJs composed of Fe(14 nm) / MgO($d$ nm) / $Ge_{0.935}Fe_{0.065}$ (50 nm) / Ge:B (B: $4\times10^{19}$ cm$^{-3}$, 70 nm) grown on a $p^+$-Ge (001) substrate by low-temperature molecular-beam epitaxy (LT-MBE) [Fig. 1(a)]. The growth process is described as follows. After the Ge substrate was chemically cleaned by ultra pure water, ammonia water, and acetone, followed by cleaning and etching with ultra pure water and buffered HF in a cyclical manner for 1 hour, it was introduced in the ultrahigh vacuum MBE growth chamber through an oil-free load-lock system. After degassing the substrate at 300°C for 30 minutes and successive thermal cleaning at 740°C for 15 minutes, we grew the Ge:B buffer layer at 300°C, which was followed by the growth of the 50-nm-thick $Ge_{0.935}Fe_{0.065}$ layer at 240°C. The MgO barrier layer was grown by electron beam deposition in our MBE growth chamber at 80°C with a growth rate of 0.02 Å/s. The thickness $d$ of the MgO barrier was changed from 3 nm to 9 nm in the same wafer by moving the main shutter in front of the sample surface during the deposition of MgO. Then, we grew the top Fe layer at 50°C. To obtain a flat surface of the top Fe layer, the sample was annealed at 250°C for 30 minutes after the growth. We used *in situ* reflection high-energy electron diffraction (RHEED) to monitor the crystallinity and surface morphology during the growth [Figs. 1(c)-1(j)]. The diffraction patterns indicate that the MgO and Fe layers are epitaxially grown on $Ge_{0.935}Fe_{0.065}$ with the epitaxial relationship of Fe[100](001)// MgO[110](001)// $Ge_{0.935}Fe_{0.065}$[100](001) shown in Fig. 1(b), which is the same as that of Fe/MgO/Ge.[14,15] The diffraction patterns of the top Fe layer change from spotty [Figs. 1(e) and 1(f)] to streaky [Figs. 1(c) and 1(d)] by the annealing, reflecting the improvement of the surface flatness. The root mean square of the surface roughness of the top Fe layer measured by atomic force microscopy (AFM) was about 0.24 nm, which means that an atomically flat surface was obtained. The $T_C$ value of the $Ge_{0.935}Fe_{0.065}$ layer was 100 K.[4] Figures 2(a) and 2(b) show the cross-sectional high-resolution transmission-electron microscopy (HRTEM) images of the sample. One can see that almost the entire region of the trilayer has an epitaxially-grown single-crystal structure with smooth and flat interfaces. There are some amorphous-like crystal domains, which are indicated by the yellow arrows in Figs. 2(a) and 2(b), between the regions that have slightly different crystal orientations indicated by the white dashed lines in Fig. 2(b).

For tunneling transport measurements, square mesas with a size of 700 × 700 μm were fabricated on the sample using photolithography and Ar-ion etching. As shown in Fig. 3(a), the resistance-area product *RA* is symmetric about *V*=0 for all the MTJs with *d* from 3 to 9 nm at 3.5 K, where *V* is the bias voltage applied to the top electrode with respect to the substrate. This suggests that the Schottky barrier is not formed at the MgO/GeFe interface. This result can be understood by considering the $E_F$ position; it was reported that $E_F$ is pinned at about 0.12 eV above the valence band maximum (VBM) at the MgO/*p*-Ge interface,[16] and recent angle-resolved photoemission spectroscopy (ARPES) measurements for GeFe showed that $E_F$ is located at 0.35 eV above the VBM in impurity bands, which are indicated by the pink area in the inset of Fig. 3(a).[9] The *RA* increases exponentially as *d* increases [Fig. 3(b)]. This is a typical feature of tunnel junctions. In the Wentzel-Kramer-Brillouin (WKB) approximation, the slope of the ln *RA* – *d* characteristics is given by $2\sqrt{2m^*V_b}/\hbar$. From our results, $m^*V_b$ [kg·eV] is estimated to be $0.035m_0$ for Fe/MgO/GeFe. Here, $m_0$ is the free-electron mass, $m^*$ is the effective mass of holes, and $V_b$ is the barrier height. This value is significantly lower than the reported values for the MgO barrier in the literature ($1.1m_0$ and $3.6m_0$ for the epitaxial MgO(001) barrier in Fe/MgO/Fe and FeCo/MgO/Fe structures, respectively).[17,18] In Fe/MgO/Fe MTJs, it is known that the barrier height is decreased by oxygen-vacancy defects in the MgO barrier.[12] The low barrier height of our junctions is probably due to the presence of the amorphous-like crystal domains in the MgO layer, which are seen in the TEM lattice images [see Figs. 2(a) and 2(b)], in addition to the oxygen-vacancy defects. These domains may have a role of leak paths and decrease the tunnel resistance, lowering the barrier height.

Figure 4(a) shows the magnetic-field dependence of *RA* measured with *V* = 40 mV at 3.5 K when *d* is 3 nm. The magnetic field *H* was applied along the [110] axis in the plane of the Ge substrate. We note that the *RA* - *H* data showed no noticeable dependence on the in-plane direction of *H*, reflecting the weak magnetic anisotropy of GeFe. The red and blue curves (major loop curves) were obtained by sweeping *H* from positive to negative and negative to positive, respectively. The jumps of *RA* at $\mu_0 H$ = ~±2 mT in the major loop correspond to the magnetization reversal of the top Fe layer.[12] The *RA* values measured with the opposite magnetic-field sweep directions gradually become closer with increasing |$\mu_0 H$|, reflecting the gradual saturation of the magnetization in the $Ge_{0.935}Fe_{0.065}$ layer.[8] As can be seen in the minor loop [green curve in the inset of Fig. 4(a)], the anti-parallel magnetization configuration is stable at $\mu_0 H$ = 0 T. This is a typical feature of TMR. We note that the measurements of *RA* - *H* performed on a reference Al/$Ge_{0.935}Fe_{0.065}$/Ge:B sample, which does not have an MgO barrier layer, did not show clear magnetoresistance, indicating that the observed magnetoresistance in Fe/MgO/GeFe originates from the tunneling transport through the MgO barrier.

We can calculate the TMR curves using the magnetization curves of Fe and GeFe. From Julliere's model,[20] the TMR ratio defined as [*RA*($\mu_0 H$) - *RA*(200 mT)]/*RA*(200 mT) is given by

$$\text{TMR ratio} = \frac{2P_{\text{Fe}}P_{\text{GeFe}}\cos(\theta_{\text{Fe}}-\theta_{\text{GeFe}})}{1-P_{\text{Fe}}P_{\text{GeFe}}\cos(\theta_{\text{Fe}}-\theta_{\text{GeFe}})}. \tag{1}$$

Here, *RA*($\mu_0 H$) represents the *RA* value obtained under $\mu_0 H$ (mT). $P_{\text{Fe}}$ ($P_{\text{GeFe}}$,) and $\theta_{\text{Fe}}$ ($\theta_{\text{GeFe}}$) are the spin polarization *P* and the direction of the magnetization relative to the [110] axis (//*H*) in the Fe (GeFe) layer, respectively. When $P_{\text{Fe}}P_{\text{GeFe}} \ll 1$, the denominator in Eq. (1) becomes 1. Because the magnetization of the Fe layer is sharply reversed (*i.e.*, $\theta_{\text{Fe}}$ = 0° or 180°), $\cos(\theta_{\text{Fe}} - \theta_{\text{GeFe}})$ is proportional to $M_{\text{Fe}}M_{\text{GeFe}}$,[19] where $M_{\text{Fe}}$ ($M_{\text{GeFe}}$) is the *H* direction component of the magnetization of Fe (GeFe). Thus, the TMR ratio of our MTJs should be approximately proportional to $M_{\text{Fe}}M_{\text{GeFe}}$. Figure 4(b) shows the calculated TMR ratio obtained by $M_{\text{Fe}}M_{\text{GeFe}}$. Here, the magnetization curve of the $Ge_{0.935}Fe_{0.065}$ layer was estimated by magnetic circular dichroism (MCD) at 5 K with a photon energy of 2.3 eV corresponding to the *L*-point energy gap of bulk Ge measured for a $Ge_{0.935}Fe_{0.065}$ film, which was grown with the same condition as that for our MTJ sample.[8] We assumed that the magnetization curve of the top Fe has a rectangular shape with the coercive fields of ±2 mT. We can see that the calculated TMR curves qualitatively reproduce the experimental TMR curves [Fig. 4(b)].

Figure 5(a) shows the TMR ratio as a function of $\mu_0 H$ at various temperatures measured with *V* = 40 mV when *d* = 3 nm. The TMR ratio (at $\mu_0 H$ = 2 mT) decreases [Fig. 5(b)] and the hysteresis becomes smaller [Fig. 5(a)] with increasing temperature, reflecting a decrease in the magnetization of the $Ge_{0.935}Fe_{0.065}$ layer.[8] These results also support our conclusion that the measured magnetoresistance is due to the TMR effect.

First-principles supercell calculations performed on GeFe suggested that the Fermi level is located in two overlapping impurity bands that have opposite spin directions formed in the band gap; one is a narrow spin-down $d(e)$ band, in which $P$ is almost 100%, and the other is a spin-up $p$-$d(t_2)$ band, in which $P$ is about 70%.[9] In the recent ARPES measurements for $Ge_{0.935}Fe_{0.065}$, it was confirmed that the Fe 3$d$ impurity states actually have finite contribution to the density of states at $E_F$.[9] Because the observed sign of the TMR ratio was positive in Fe/MgO/GeFe and $P$ of Fe is positive, our result means that the spin-up $p$-$d(t_2)$ band is responsible for the tunneling properties.[9] From the TMR ratio about 0.27% observed at 3.5 K [Fig. 4(a)], the $P$ value of the $Ge_{0.935}Fe_{0.065}$ layer is estimated to be 0.17% by Julliere's model,[20] when we use the effective $P$ value of the top Fe layer estimated from TMR observed for Fe/MgO/Fe ($P$ = 75%).[12] This $P$ value obtained for GeFe is much smaller than that predicted by the first-principles supercell calculations (70%).[9] This is probably due to the leak current, which does not contribute to TMR, through the amorphous crystal domains. Thus, the TMR ratio is expected to be enhanced in the Fe/MgO/GeFe MTJs by improving the crystallinity of the MgO tunnel barrier and decreasing the leak current through the amorphous crystal domains in MgO.

In summary, we have grown MTJs composed of epitaxial Fe/MgO/$Ge_{0.935}Fe_{0.065}$ and demonstrated the first successful observation of TMR in the MTJs containing a group-IV ferromagnetic semiconductor. This result confirms the presence of spin-polarized carriers at $E_F$ in GeFe. The observed sign of the TMR was positive, which revealed that the largely spin-polarized carriers in the $p$-$d(t_2)$ band are dominant for the tunneling. The TMR ratio will be increased in the Fe/MgO/GeFe MTJs by improving the crystallinity of the MgO tunnel barrier and decreasing the leak current through the amorphous crystal domains in MgO. Our results show that GeFe is promising for spin injectors and detectors of future Si- and Ge-based spintronic devices.

**ACKNOWLEDGEMENTS**
This work was partly supported by Grants-in-Aid for Scientific Research (23000010, 26249039) including the Specially Promoted Research, Project for Developing Innovation Systems from MEXT, Spintronics Research Network of Japan (Spin-RNJ), and the Cooperative Research Project Program of RIEC, Tohoku University. This work was partly conducted in Research Hub for Advanced Nano Characterization. Y. K. W. acknowledges financial support from JSPS through the Program for Leading Graduate Schools (MERIT) and the JSPS Research Fellowship Program for Young Scientists.

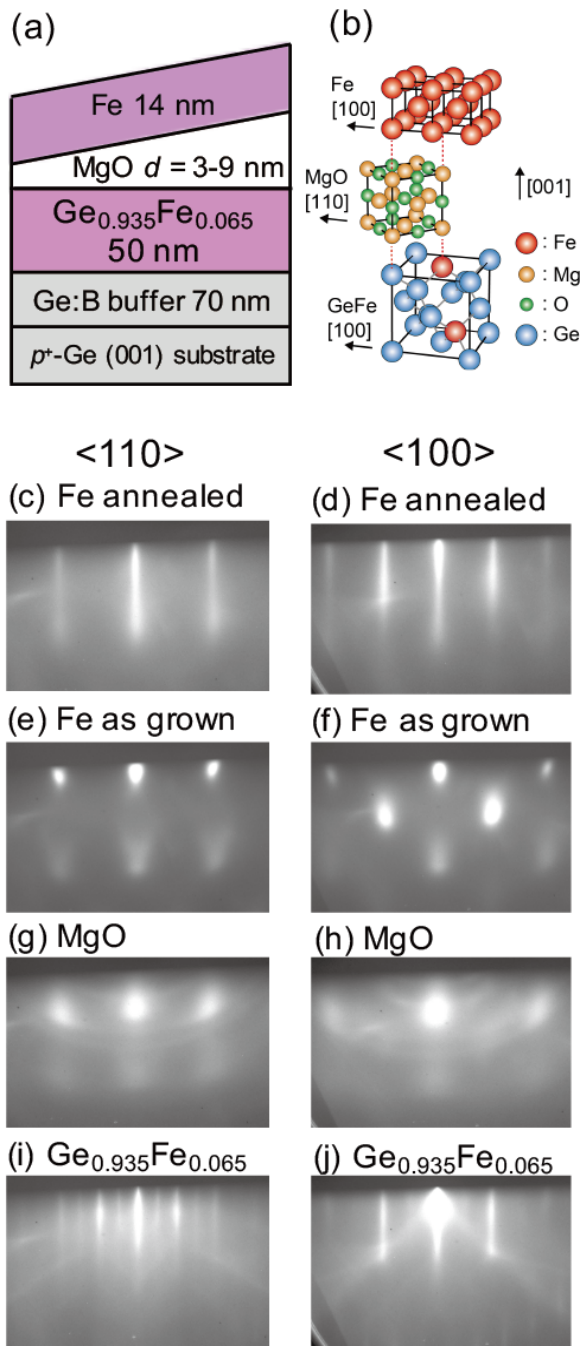

Fig. 1. (a),(b) Schematic structures of (a) the Fe/MgO/GeFe trilayer sample fabricated in our study and (b) its epitaxial relationship. In (b), the black and gray lines represent the unit cells and covalent bonds, respectively. (c)-(j) RHEED patterns of (c),(d) the annealed Fe layer, (e),(f) as-grown Fe layer, (g),(h) MgO layer, and (i),(j) $Ge_{0.935}Fe_{0.065}$ layer with the electron-beam azimuth along (c),(e),(g),(i) the <110> direction and (d),(f),(h),(j) the <100> direction of the Ge(001) substrate.

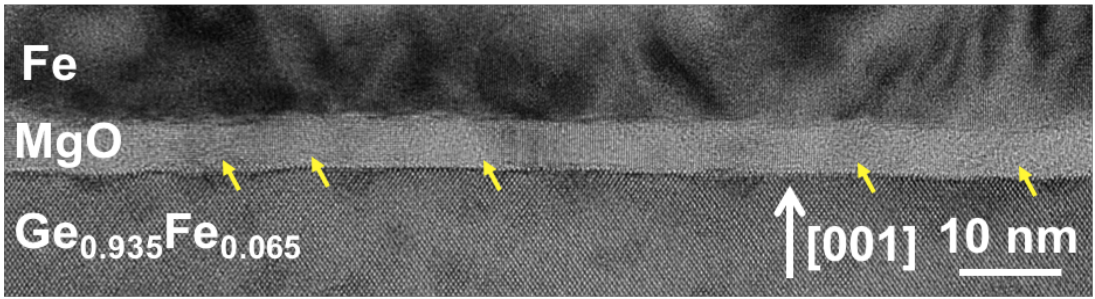

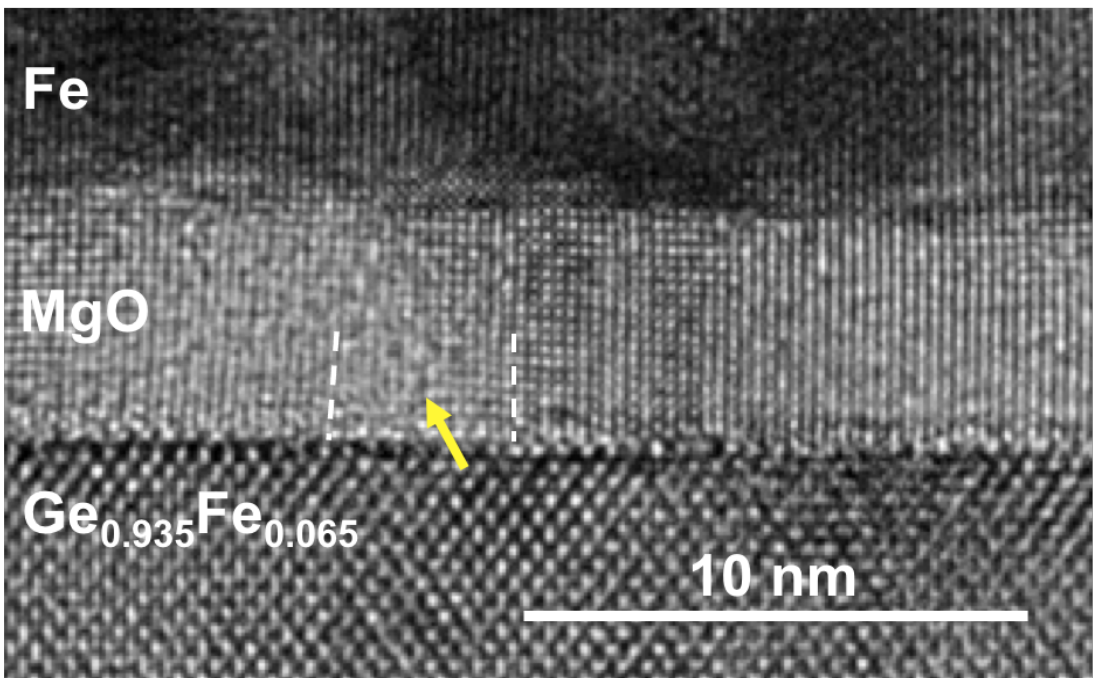

Fig. 2. (a) HRTEM lattice image of the Fe/MgO/GeFe sample projected along the Ge<110> axis. (b) Magnified image of (a). In the MgO layer, there are some amorphous-like crystal domains indicated by the yellow arrows between the regions that have slightly different crystal orientations indicated by the white dashed lines.

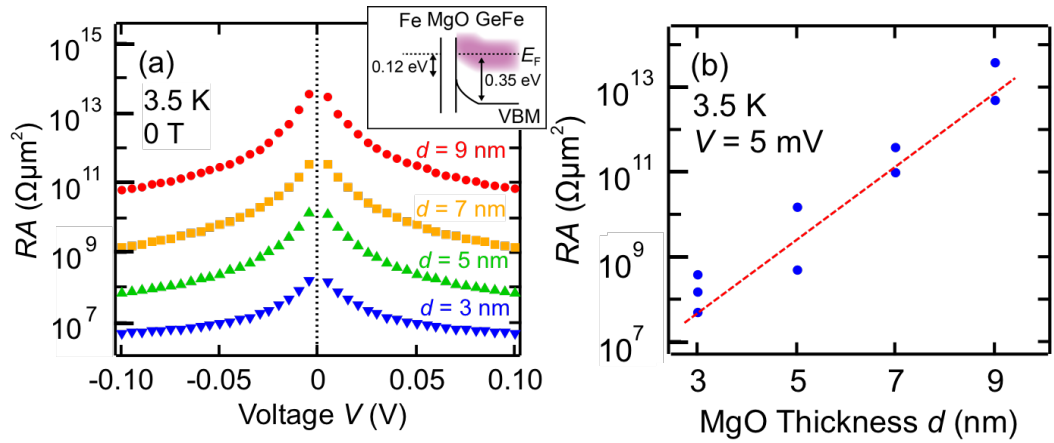

Fig. 3. (a) *RA* versus the bias voltage *V* measured at 3.5 K for the Fe/MgO/GeFe MTJs with the MgO thickness *d* ranging from 3 to 9 nm. Inset of (a) shows the band line-up of Fe/MgO/Ge$_{0.935}$Fe$_{0.065}$. The solid and dotted lines correspond to the VBM and the Fermi level $E_F$. The pink area represents the impurity bands. (b) *RA* as a function of *d* measured with *V* = 5 mV at 3.5 K.

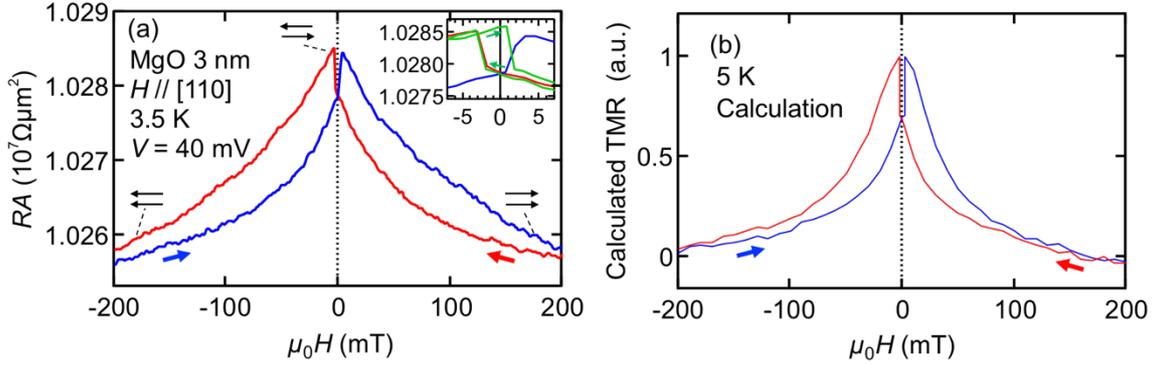

Fig. 4. (a) Magnetic-field ($\mu_0 H$) dependence of $RA$ of the MTJ with $d$ = 3 nm measured with a bias voltage $V$ of 40 mV at 3.5 K. The magnetic field $H$ was applied along the [110] axis in the plane of the Ge substrate. The red and blue curves (major loop curves) were obtained by sweeping $H$ from positive to negative and negative to positive, respectively. The black arrows indicate the magnetization configurations of the top Fe layer and the bottom $Ge_{0.935}Fe_{0.065}$ layer. The inset shows the magnified plot of the TMR curves. The green curve is the minor loop. (b) Calculated TMR curves at 5 K as a function of $\mu_0 H$ obtained by multiplying the [110] direction component of the magnetizations between the top Fe and the bottom $Ge_{0.935}Fe_{0.065}$ layers.

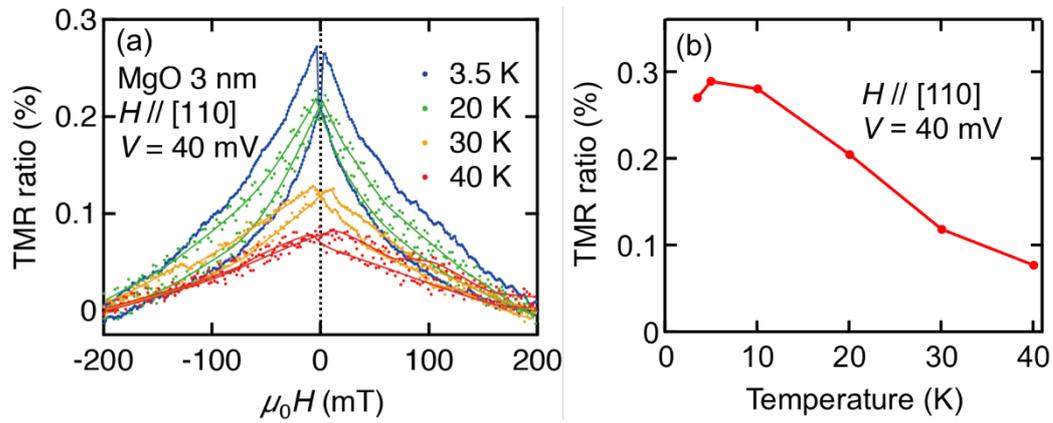

Fig. 5. (a) TMR ratio, which is defined as [$RA(\mu_0 H)$ - $RA$(200 mT)]/$RA$(200 mT), as a function of $\mu_0 H$ at various temperatures. (b) TMR ratio at $\mu_0 H$ = 2 mT as a function of temperature when $d$ is 3 nm and the bias voltage $V$ is 40 mV.